\documentclass[slac_one]{revtex4}
\usepackage{graphicx}
\usepackage{fancyhdr}
\usepackage{wrapfig,rotating}
\def\met{\mathbin{E\mkern - 11mu/_T}}

\begin{document}

\title{Vector Boson + Heavy Flavor Jets Production at the Tevatron} 

%

\author{Kenichi Hatakeyama}
\affiliation{Rockefeller University, New York, NY 10065, USA.\\
for the CDF and D0 Collaborations}
%

\begin{abstract}
Recent measurements on the vector boson plus
heavy-flavor jets production by the CDF and D0 experiments
are presented in comparisons with recent theoretical predictions.
Good understanding of such processes is important to improve our
understanding of QCD and also to enhance the potential to search for
yet-to-be-discovered new physics phenomena which lead to similar
final states.
\end{abstract}

\maketitle

\thispagestyle{fancy}

\section{Introduction}

Events containing vector bosons and heavy-flavor jets
constitute valuable data samples for a variety of
physics analyses at both CDF and D0 experiments
at the Fermilab Tevatron $p\bar p$ collider.
For example, an event sample containing a $W/Z$ boson
and $b$-jet(s) is a golden sample for the Higgs search at low masses,
studies of top quark production and properties, 
and also searches for the beyond-the-standard-model phenomena.
Thus, the measurements of cross sections for QCD production of vector
boson + heavy-flavor jets are important not only to enhance our
understanding of QCD but also to better understand the backgrounds to 
other important physics studies.
Presented below are recent measurements by the CDF and D0 experiments
on these processes.


\section{W+b-Jets Production}

\begin{wrapfigure}{r}{0.39\columnwidth}
\centerline{
\includegraphics[width=0.39\columnwidth,bb=35 35 520 495,clip=]
{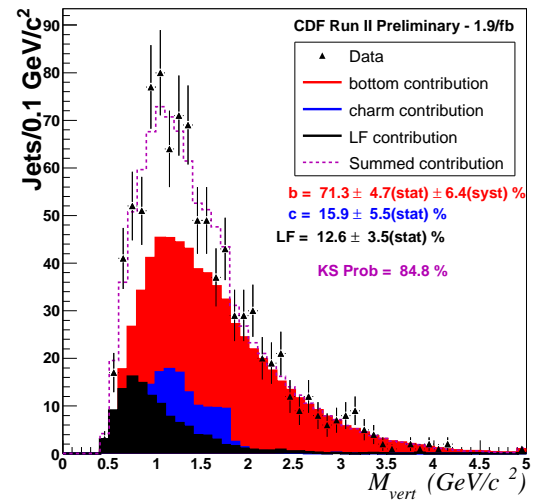}
}
\caption{Secondary vertex mass distribution for $b$-tagged jets
in the selected data sample.}\label{fig:Wb}
\end{wrapfigure}

CDF has made a measurement of the $W+b$-jets cross section~\cite{Wb_CDF}
using 1.9 fb$^{-1}$ of $p\bar p$ collision data at 
$\sqrt{s}=1.96$ TeV.
The event sample used in this measurement is selected from the
high $p_T$ electron and muon samples
by requiring a reconstructed isolated
electron with $E_T>20$ GeV or muon with $p_T>20$ GeV/{\it c},
missing transverse energy ($\met$) $>25$ GeV,
and one or two jets with $E_T>20$ GeV and $|\eta|<2$.

$b$-jets are ``tagged'' by the presence of a displaced secondary
vertex within the jet arising from the decay of the long-lived $B$
hadron.
The $b$-tagged jet sample includes background contributions from
charm and light-flavor (LF) jets. The fraction of $b$-jets in the
$b$-tagged sample is estimated by performing a fit to the invariant
mass of all charged tracks associated with the secondary vertex as shown in
Fig.~\ref{fig:Wb}; on average,
$b$-jets have a larger mass than $c$-jets and LF jets due to
the hierarchy of $B$, $D$ and LF hadron masses.

There are non-negligible contributions to $W+b$-jet
candidates from processes other than QCD $W+b$ production.
The largest background contribution comes from QCD multijet production
in which there is no real $W$, but a combination of jets faking
leptons, mismeasured jet energy, or semileptonic $b$-decay make the
events pass through all the event selections.
The other major background sources are $t\bar t$ production and single
top quark production. The QCD multijet production contribution is
estimated using data, and the other background contributions are
estimated from MC predictions.

The cross section for $b$-jets with $E_T>20$ GeV and $|\eta|<2$ 
from
%
QCD $W+b$ production is measured to be
$\sigma(Wb)\times BR(W\to l^{\pm}\nu)
=2.74 \pm 0.27\mbox{(stat)} \pm0.42\mbox{(syst)} \mbox{~pb}$
which is higher than the Alpgen prediction of 0.78 pb by $\sim 3.5$.
This large disagreement is somewhat unexpected and indicates a need
for an improved theoretical prediction for this process.

\section{Z+b-Jets Production}

The $Z+b$-jets production has been studied by both D0~\cite{Zb_Dzero} 
and CDF~\cite{Zb_CDF}.
The dominant production diagrams contributing to the $Z+b$-jets
final state are (a) $bg\to Zb$ ($\sim65$\%) and 
(b) $q\bar q\to Zb\bar b$ ($\sim35$\%) in
next-to-leading order (NLO) pQCD predictions.
The cross section is sensitive to the $b$-quark density in
the proton, and thus the cross section measurement 
provides important information for the $b$-quark density
which is so far indirectly extracted from gluon
density.
A good understanding of the $b$ density is essential
to accurately predict the production of particles that couple strongly to
$b$-quarks including supersymmetric Higgs bosons and single top
production.
$Z+b$-jets production is also a major background in searches for the
Higgs production in the $ZH\to Zb\bar b$ channel.

\begin{figure}[tp]\centering\leavevmode
  \begin{tabular}{cc}
  \includegraphics[width=0.45\hsize,clip=]
  {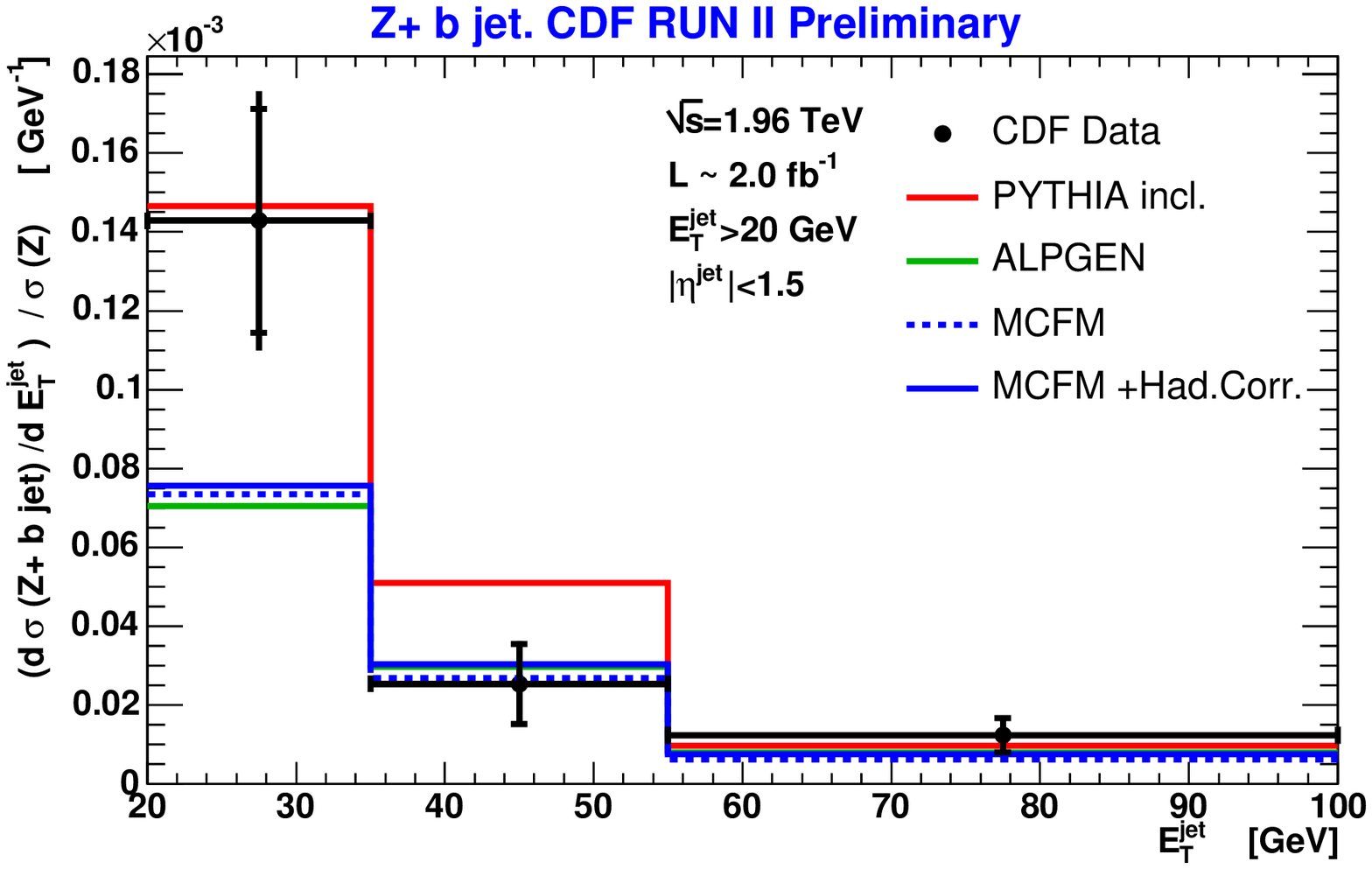}&
  \includegraphics[width=0.45\hsize,clip=]
  {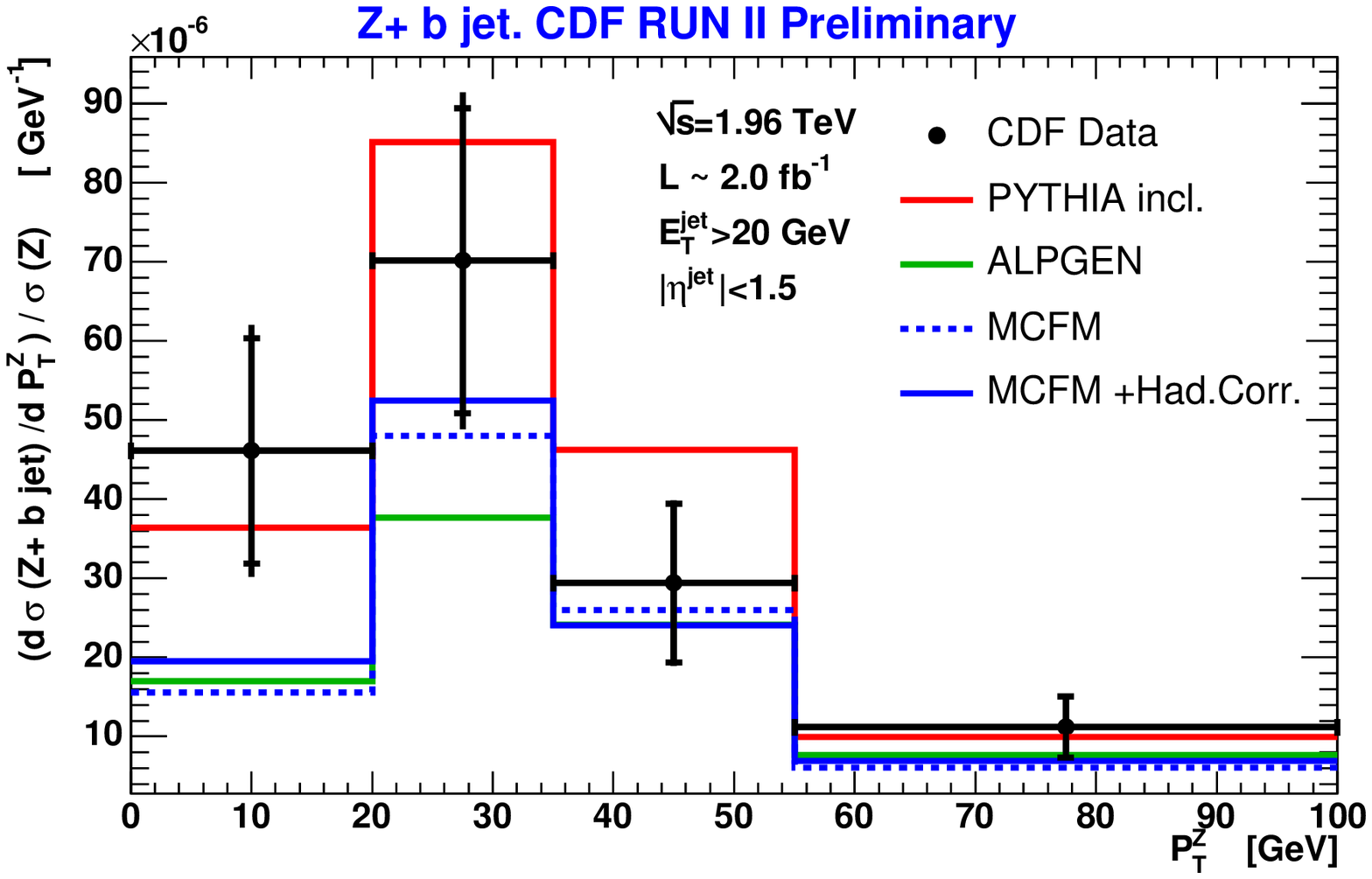}
  \end{tabular}
  \caption{$Z+b$-jets differential cross sections as a function of jet
    $p_T$ (left) and $Z$ $p_T$ (right) from the CDF 2~$\mbox{fb}^{-1}$
    of data.}
  \label{fig:Zb}
\end{figure}

CDF has recently updated the measurement on $Z+b$-jet production
using 2.0 fb$^{-1}$ of data.
The measurement was made using jets with $E_T>20$ GeV and $|\eta|<1.5$
tagged as $b$-jets by the secondary vertex algorithm in $Z\to e^+e^-$
and $Z\to\mu^{+}\mu^{-}$ events
and the results are summarized in Table.~\ref{tab:zb}.
%
%
The high statistics in the data used in the recent analysis
also allowed the first measurement of differential distributions,
which are shown in Fig.~\ref{fig:Zb} together with several
theoretical predictions.
The data are in general agreement with the data, but differences at
the level of up to 2$\sigma$ are observed.
Both Alpgen and MCFM NLO predictions lie somewhat below the data at
low jet $E_T$ ($Z$ $p_T$), but agree better with data at higher jet
$E_T$ ($Z$ $p_T$).
Pythia is in good agreement with data at low jet $E_T$, but less so at
higher jet $E_T$.
Large variations in theoretical predictions are not well understood
and need to be resolved.

\begin{table}
\caption{Results on the $Z+b$-jets production.}
{\small
\begin{tabular}{cccccc}
\hline\hline
 & CDF Data &~~Pythia~~&~~Alpgen~~&~~MCFM NLO~~& ~~MCFM NLO+UE~~       \\
 &          &          &          &      & ~~+Hadronization~~\\
\hline
$\sigma(Z+b\mbox{-jet})$      &
~~$0.86\pm0.14\pm0.12$ pb~~   &
  $-$      &   $-$    & ~~0.51~pb~~  &  0.53~pb    \\
$\sigma(Z+b\mbox{-jet})/\sigma(Z)$       &
  $0.336\pm0.053\pm0.041$\%   &
  0.35\%  &  0.21\% &  0.21\% &  0.23\%   \\
~~~$\sigma(Z+b\mbox{-jet})/\sigma(Z+\mbox{jet})$~~~ &
  $2.11\pm0.33\pm0.34$ \%     &
  2.18\%  &  1.45\% &  1.88\% &  1.77\%   \\
\hline\hline
\end{tabular}
}
\label{tab:zb}
\end{table}

\section{Photon+Heavy-Flavor Jet Production}

In the standard model (SM), $\gamma+b/c$ events are
produced predominantly produced via the compton scattering
$Q+g\to Q+\gamma$ (where $Q=b/c$) at low $p_T$,
and the contribution of the annihilation process $q\bar q
\to Q\bar Q\gamma$ increases with increasing $p_T$.
The cross section is sensitive to the $b$-quark density in
the proton as in the case of the $Z+b$-jet production.
The good understanding of these processes is also important as
the QCD production of $\gamma+b/c$ is a significant background
in new physics searches, including searches for
the techniomega production
($\omega_{TC}\to\gamma\pi_{TC}\to \gamma b\bar b$),
some SUSY scenarios, and excited $b$-quark production.

The $\gamma+b$-jet cross section was measured by
CDF~\cite{gammab_CDF} and D0 reported improved measurements on
$\gamma+b/c$-jet recently~\cite{gammaHF_D0} based on the
1 $\mbox{fb}^{-1}$ of data.
The measurement is made in the kinematic region of
$30<p_T^{\gamma}<150$ GeV/{\it c},
$|y^{\gamma}|<1$,
$p_T^{jet}>15$ GeV/{\it c}, and
$|y^{jet}|<0.8$
for events with $y^{\gamma}y^{jet}>0$ (region 1) and
$y^{\gamma}y^{jet}<0$ (region 2), separately.
These two rapidity combinations helps to differentiate the parton $x$
regions contirbuting to the two regions. The regions 1 and 2 are
sensitive to $0.01<x_1<0.03$, $0.03<x_2<0.09$, and $0.02<x_1,x_2<0.06$,
respetively.

Photon candidates are selected using a neural network (NN)
approach~\cite{gammajet_D0}.
The photon purity is higher than $50$\% in all kinematic regions and
improves with increasing $p_T^{\gamma}$.
The heavy-flavor jet tagging is also performed using a NN which
exploits the longer lifetime of $B/D$ hadrons compared to the lighter
ones. 
The inputs to the neutral network include the number of
secondary vertices, secondary vertex mass, a weighted combinations of
the track's impact parameter significances and the probability that a
jet originates from the primary vertex (JLIP probability).
Among the tagged jets,
the fraction of $b$, $c$, and light-flavor jets is determined based on
a template fit to the JLIP probability distribution.

\begin{figure}[tp]\centering\leavevmode
  \begin{tabular}{c}
  \includegraphics[bb=0 25 567 505,width=0.42\hsize]
  {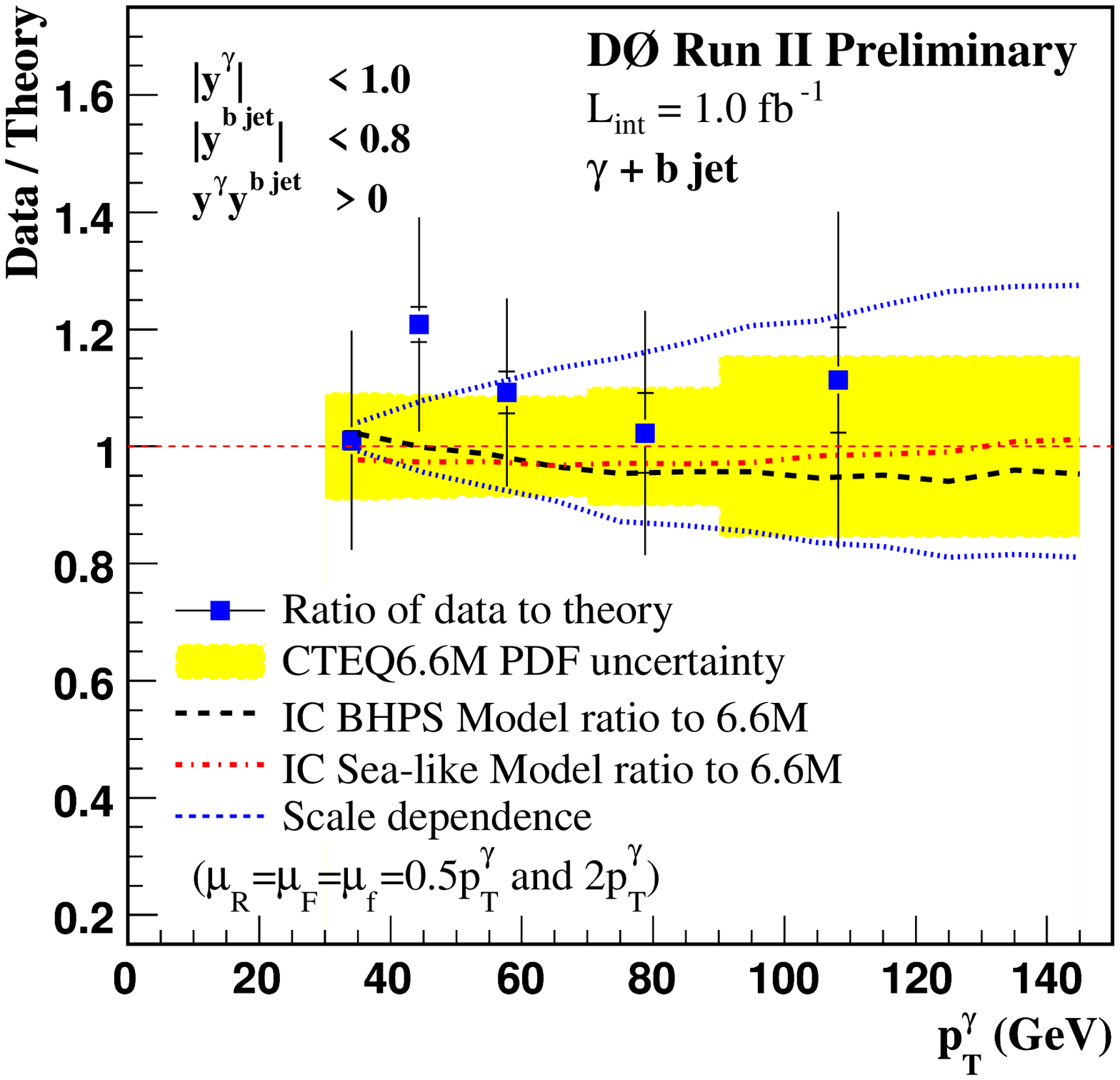}
  \includegraphics[bb=0 25 567 505,width=0.42\hsize]
  {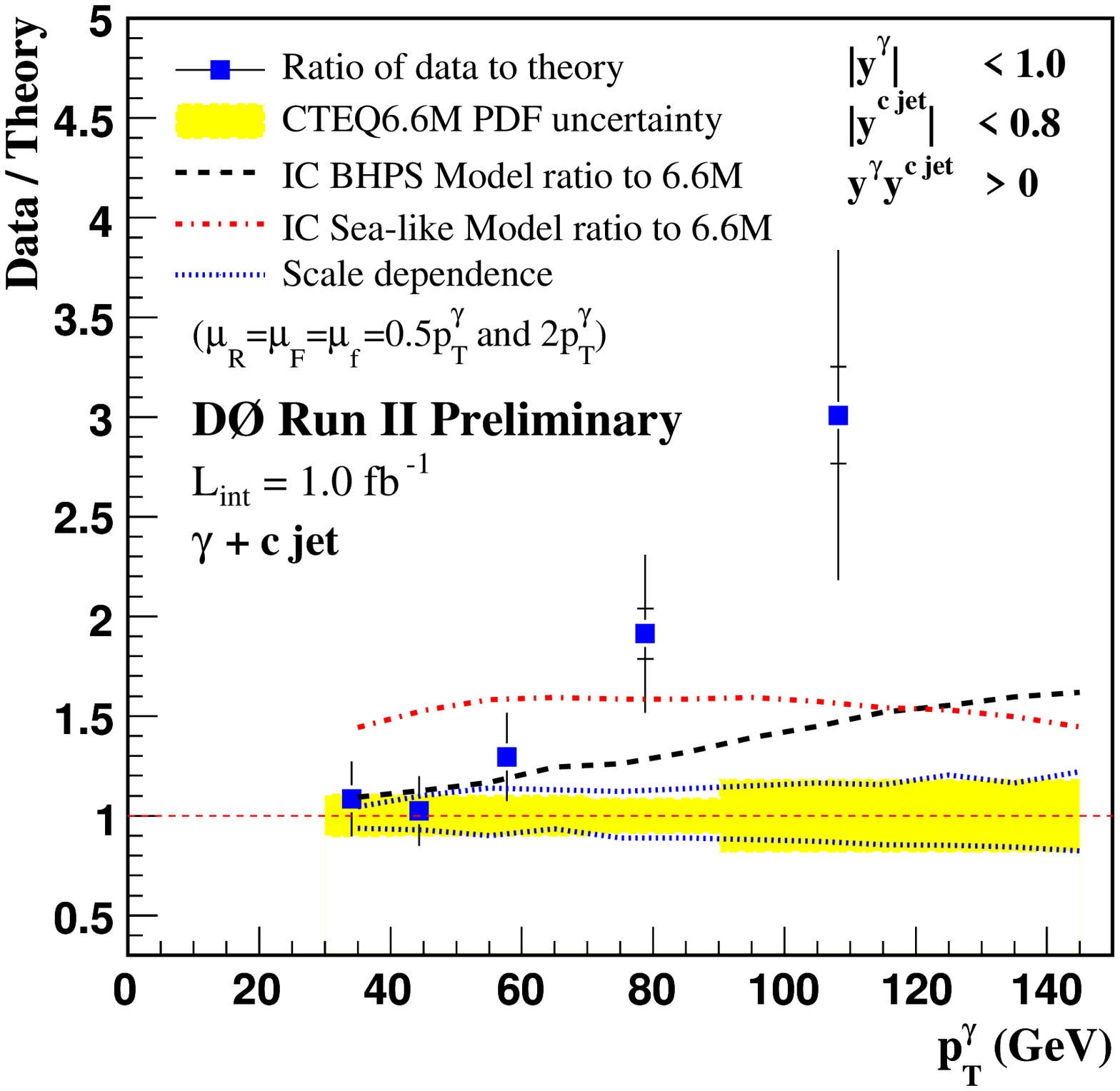}
  \end{tabular}
  \caption{The $\gamma+b$-jet (left) and $\gamma+c$-jet cross section
    ratio of data to theory as a function of $p_T^{\gamma}$ in the
    region 1 from D0.}
  \label{fig:photonHF}
\end{figure}

The measured differential cross sections for the $\gamma+b/c$-jet
production are compared to their theoretical predictions from NLO pQCD
as functions of $p_T^{\gamma}$ in Fig.~\ref{fig:photonHF}.
For $\gamma+b$, data and theoretical predictions are in good agreement
over the full kinematic region explored. 
For $\gamma+c$, a reasonable agreement is observed only at
$p_T^{\gamma}<50$ GeV/c, and the deviation increases with increasing
$p_T^{\gamma}$ in both regions 1 and 2.
The deviation may be attributed to a possible non-negligible intrinsic
charm content in the proton and/or the inaccurate description of
$g\to c\bar c$ fragmentation.

\section{W+Single Charm Production}

$W+c$ events are produced
by $gs\to Wc$ ($\sim90$\%) and $gd\to Wc$ ($\sim10$\%)
in the SM.
Thus, the production cross section is sensitive to the
$s$-quark PDFs in the proton at a scale on the order of the
$W$ mass, and it is also sensitive to the element of the
CKM matrix $V_{cs}$.
Also, $Wc$ is an important component of the W+1 and 2 jet
samples, that are used in searches for {\it e.g.} a single
top, the Higgs boson, and a supersymmetric top,
and the searches will benefit from good understanding
of QCD $Wc$ production.

Both CDF~\cite{Wc_CDF} and D0~\cite{Wc_Dzero} studied W+single $c$
production recently.
In both measurements,
$W\to l\nu$ events are selected by requiring a high
$p_T$ isolated electron or muon with large $\met$.
Charm jets are identified from their semileptonic decay
by looking for a muon within the jet; this charm jet identification
algorithm is referred to as the soft lepton tagging (SLT) algorithm.
$W+c$ events are identified by utilizing the charge correlation
between a lepton from $W$ decay ($W$ lepton) and SLT muon,
{\it i.e.}, the difference between events in which the $W$ lepton
and SLT muon have opposite charge (OS) and events in which they have
same charge (SS).
The $Wc$ production mainly leads to OS events; however most of background
processes such as the $Wc\bar c$ production give OS and SS events
almost equally.
Therefore, events from $Wc$ production can be extracted from the
excess of OS-SS events.

\begin{wrapfigure}{r}{0.40\columnwidth}
\centerline{\includegraphics[bb=0 0 275 292,width=0.32\columnwidth,clip=]
  {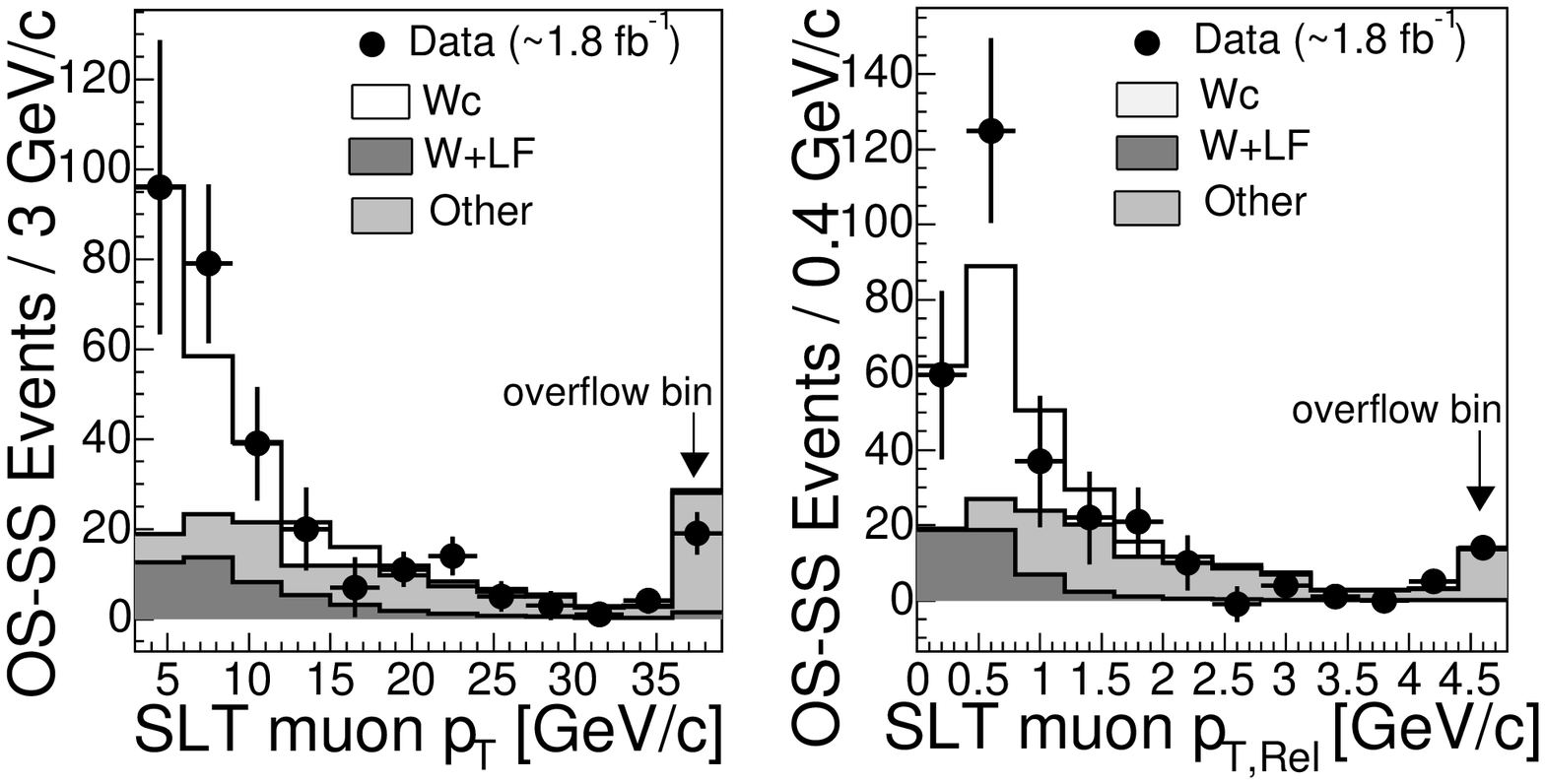}}
\centerline{\includegraphics[bb=0 14 567 500,width=0.40\columnwidth,clip=]
  {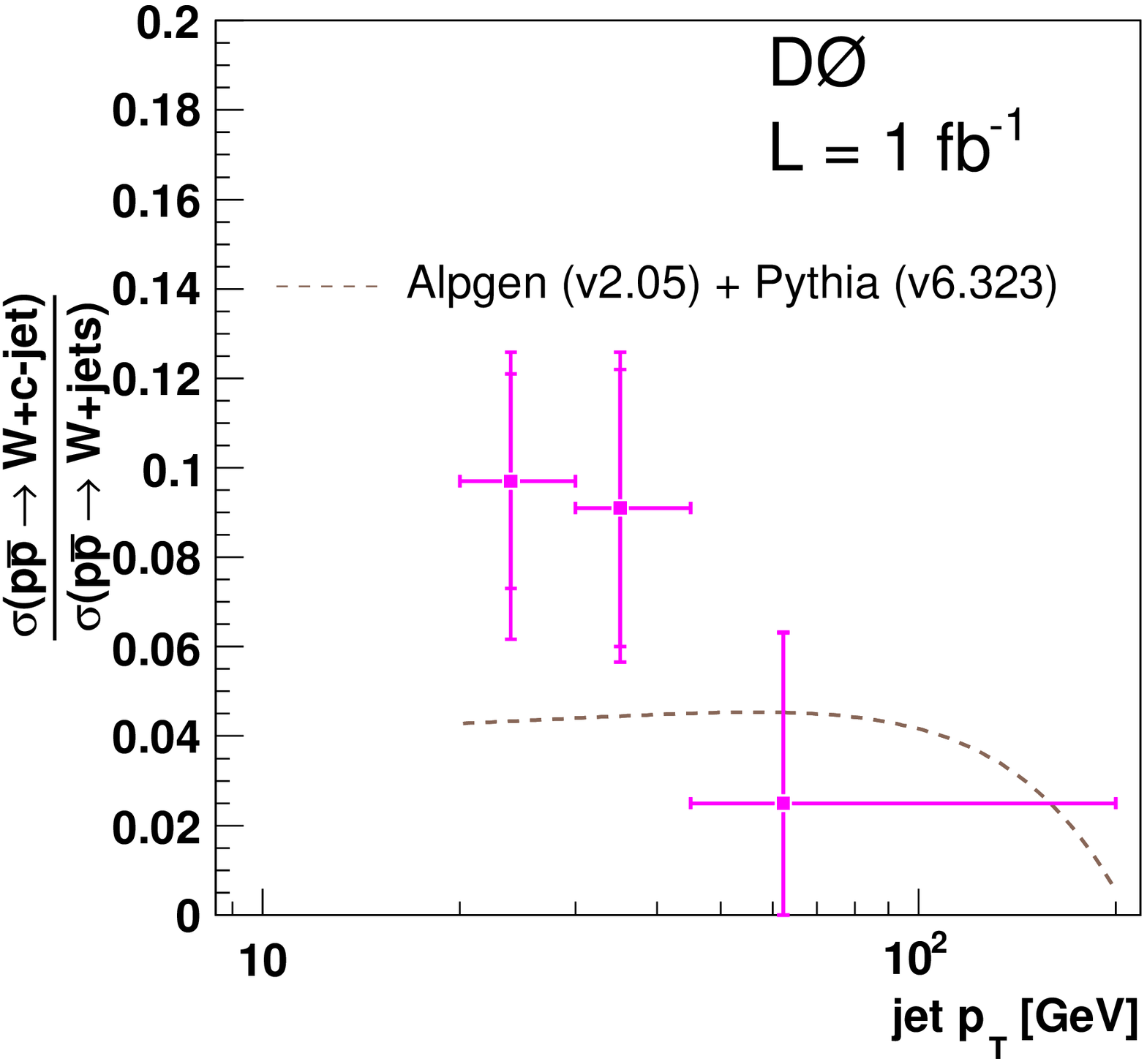}}
  \caption{(top) OS-SS events 
    as a function of SLT muon $p_T$ measured by CDF.
    (bottom) Ratio $\sigma(W+c\mbox{-jet})/\sigma(W+\mbox{jets})$
    for jet $p_T>20$ GeV and $|\eta|<2.5$ measured by D0.}
  \label{fig:Wc}
  \vspace*{-1.5cm}
\end{wrapfigure}

Figure~\ref{fig:Wc} (top) shows
the number of OS-SS events as a function of SLT muon $p_T$ measured by
CDF, with an excess which is consistent with the presence of the $Wc$
production.
After taking into account the OS-SS events from backgronds, such as
$W$+light-flavor jets, Drell-Yan, and QCD multijet 
production, 
the cross section of the $Wc$ production for $p_{T}^{c}>20$
GeV/{\it c} and $|\eta^{c}|<1.5$ is measured to be
$\sigma(Wc)\times BR(W\to l\nu)
=9.8\pm3.2$ pb
which is in good agreement with the NLO pQCD prediction of
$11.0^{+1.4}_{-3.0}$ pb.

D0 measured the cross section ratio
$\sigma(W+c)/\sigma(W+\mbox{jets})$,
since in the ratio many systematic uncertainties cancel.
The measurment was made as a function of jet $p_T$ as shown
in Fig.~\ref{fig:Wc} (bottom).
The cross section ratio integrated over $p_T^{jet}>20$ GeV/{\it c}
and $|\eta|<2.5$ is measured to be
$0.074\pm0.019(\mbox{stat})^{+0.012}_{-0.014}$
which is somewhat higher but consistent with the Alpgen prediction of
$0.044\pm0.003$.


\section{Summary}
Final states containing a vector boson and heavy-flavor jets appear
in many interesting physics processes. 
The good understanding of QCD production of such final states
is critical for physics analyses at the Tevatron and
also at the upcoming LHC, and
both CDF and D0 Collaborations have made extensitive studies on these
processes.
It was found that $W+c$ and $\gamma+b$ measurements are well
described by the state-of-the-art recent theoretical calculations;
however, the measurements on the $W/Z+b$-jets and $\gamma+c$-jet
indicate a need of an improved understanding of these processes.
The measurements with improved precision from the Tevatron experiments
would be the keys for deeper understanding of these processes and
benefit for future physics analyses at the Tevatron and also at the
LHC.

\end{document}